\documentstyle[aps,prl]{revtex}

\newcommand{\kbar}{k\hspace{-0.45em}\raisebox{0.7ex}{-}\hspace{0.2em}}
\def\be{\begin{eqnarray}}
\def\ee{\end{eqnarray}}

\def\ket#1{| #1 \rangle}

\begin{document}
%\twocolumn
\title{Dynamical Localization in the Paul Trap}
\author{M.\ El Ghafar, P.\ T\"orm\"a, V.\ Savichev, E.\ Mayr, 
A.\ Zeiler and W.\ P.\ Schleich\\
%EndAName
Abteilung f\"ur Quantenphysik, Universit\"at Ulm, 89069 Ulm, Germany}
\date{}
\maketitle

\begin{abstract}
We show that quantum localization occurs in the center-of-mass motion
of an ion stored in a Paul trap and interacting with a standing laser field.
The present experimental state of the art makes the observation of this
phenomenon feasible.
\end{abstract}
%\narrowtext
The phenomenon of dynamical localization -- an analogue of
Anderson localization \cite{Anderson} of electronic waves in one-dimensional
disordered solids
is a finger print of quantum chaos \cite{Haake92}. This has motivated
the experimental verification of
the suppression of ionization of Rydberg atoms in
microwave fields~\cite{review} and the localization in the momentum
distribution of an atom moving in a phase modulated standing wave
\cite{Moore,Bialynicki-Birula94}. In this letter we show that the effect
of dynamical localization of quantum mechanical wave packets 
appears also in the center-of-mass motion of a single ion 
confined in a Paul trap \cite{Paul} and interacting with a
laser field. This system is of particular interest, since
here the spatial periodicity of the standing wave is broken by the
binding potential of the trap. This results in the fact that
in contrast to the previous examples we find localization
both in the momentum and the position variables. Moreover
we find a characteristic three-peak structure in the quantum
mechanical position distribution. This structure is a pure quantum 
effect since it is completely absent in the classical position
distribution. The center peak at the origin is the remnants of the
initial distribution, whereas the side peaks may be attributed to
quantum tunneling. The recent experimental successes
\cite{Wineland} in controlling the quantum motion in a Paul trap
make our proposal experimentally feasible.

The phenomenon of dynamical localization in the
Paul trap emerges because  
(i) the Paul trap is an explicitely time dependent
device, (ii) a standing laser wave provides a spatially periodic light 
potential for the center-of-mass motion, and (iii) the temperature of the
ion is so low that its motion
has to be treated quantum mechanically \cite{Blumel}.
Starting from the Hamiltonian
describing the motion in the Paul trap in the presence of
the standing wave we compute the position and momentum distributions of the 
ion by solving the corresponding Newton's equations and 
the Schr\"odinger equation. 
We show that the classical distributions are broad Gaussians \cite{Chacon94}. 
In contrast, the quantum distributions display 
on top of a broad background a narrow three-peaked
distribution. To identify the origin of this
effect we calculate the Floquet states of this system. We show that
indeed only few Floquet states contribute significantly to the time
evolution of an initially localized wave packet in the Paul trap. We 
conclude by discussing experimental possibilities for observing this
phenomenon. 

We consider the standard Paul trap set-up realized experimentally 
in many labs \cite{Wineland,Walther,Toschek}: a standing laser 
field of frequency $\omega_L$ and wave vector $k$
aligned along the $x$-axis couples the internal states of a single 
two-level ion of mass $m$ to the center-of-mass motion. The resulting dynamics 
of the state vector $\ket{\Psi(\tilde{t})}$ 
describing the internal and external states of the ion follows from the
Schr\"odinger equation
$i \hbar \frac{\partial}{\partial \tilde{t}}
 \ket{\Psi(\tilde{t})}  = \hat{\tilde{H}} \ket{\Psi(\tilde{t})}$,
with the Hamiltonian
\begin{eqnarray}
\hat{\tilde{H}}& = & \frac{\hat{\tilde{p}}^2}{2m}+\frac{1}{2} 
\frac{m\omega^2}{4}[a+2q\cos\left(\omega \tilde{t}\right)]
\hat{\tilde{x}}^2 \nonumber \\
 && +\frac{1}{2}\hbar\omega_a \hat{\sigma}_z + \hbar\Omega_0 
\hat{\sigma}_x \cos(k\hat{\tilde{x}} +\phi)
\cos(\omega_L \tilde{t})  .
\label{1}
\end{eqnarray}
Here the parameters $a$ and $q$ denote \cite{Paul}
the DC and AC voltages applied to the trap.
The frequency of the AC field is $\omega$. The term proportional to the
Pauli spin matrix $\hat{\sigma}_z$ takes into account the internal
states of the ion with the transition frequency $\omega_a$ 
and the Pauli spin matrix $\hat{\sigma}_x$
describes the interaction with the standing wave.
Here $\Omega_0$ is the Rabi-frequency
and $\phi$ is the phase of the standing wave.

The phenomenon of dynamical localization is a
quantum coherence effect \cite{Haake92}. 
It is therefore extremely sensitive \cite{11} to 
noise such as spontaneous emission. 
In order to avoid spontaneous emission we consider 
the ion to be initially in its internal ground
state and the laser field to be strongly detuned. In the
rotating wave approximation the Hamiltonian (\ref{1}) reduces 
\cite{Kazantsev90} then to 
\begin{eqnarray}
\hat{\tilde{H}} = \frac{\hat{\tilde{p}}^2}{2m} + 
\frac{m\omega^2}{8} [a + 2 q \cos (\omega \tilde{t})] \hat{\tilde{x}}^2  
+ \frac{\hbar\Omega_0^2}{8\Delta} \cos [2(k\hat{\tilde{x}}+\phi)]  
\nonumber
\end{eqnarray}
where $\Delta = \omega_L - \omega_a$ is the detuning parameter. Here we have 
neglected constant energy terms. 

When we now introduce the dimensionless position $x\equiv 2 k \tilde{x}$,
time $t\equiv \omega \tilde{t}/2$ and momentum $p\equiv
\frac{4 k}{m \omega} \tilde{p} $ 
the dimensionless Hamiltonian 
\begin{eqnarray} 
\hat{H} \equiv \frac{16k^2}{m\omega^2} \hat{\tilde{H}} = 
\frac{1}{2} \hat{p}^2+\frac{1}{2}(a+2q
\cos  2 t) \hat{x}^2 + \Omega\cos (\hat{x}+2\phi)  \nonumber
\end{eqnarray}
with the effective coupling constant 
$\Omega=\frac{2\hbar k^2\Omega_0^2}{m\omega^2\Delta}$
governs via the Schr\"odinger equation
\begin{equation}
i \kbar \frac{\partial}{\partial t} \ket{\psi (x,t)} 
= \hat{H} \ket{\psi (x,t)} \label{t_ev} 
\end{equation}
the vibratory motion of the ion described by the state $\ket{\psi(x,t)}$.
Here the effective Planck constant $\kbar=\frac{8 k^2 \hbar}{m \omega}$ 
follows from the commutation relation $[\hat{x},\hat{p}] = 
\frac{8 k^2}{m \omega}[\hat{\tilde{x}},\hat{\tilde{p}}] =
\frac{8 k^2}{m \omega} i \hbar = i \kbar$.

Dynamical localization arises from the properties of the 
quantum evolution in the domain of classically chaotic dynamics. We therefore
choose the phase $\phi$ as to obtain a maximally chaotic regime. In
the following we consider the situation where the 
maximum of the cosine potential
is located at the center of the trap potential, 
i.e.~$\phi=0$ when $\Delta > 0$ or $\phi=\pi/2$ when $\Delta<0$.
This case corresponds to the appearance of 
an effective potential barrier at the center of the trap potential. 
With growing coupling constant $\Omega$, the 
unstable motion near the top of the barrier gives 
rise to the stochastic mixing of
the different classical trajectories.
In classical dynamics this leads into chaos \cite{phi}.

In order to show that indeed the classical dynamics is chaotic, we plot in
Fig.1 the Poincar\'e surface of section for the trap parameters $a=0$,
$q=0.4$ selected from the stable region of the Mathieu equation
and the coupling $\Omega=0.65$. We observe a chaotic sea with two 
stable islands in the neighbourhood of the minima 
of the standing laser field at $x=\pm\pi$. Those regular structures are 
remnants of the integrable cases $\Omega=0$ corresponding to the Mathieu
equation and the driven pendulum when $a=q=0$.

We now calculate the time-evolution 
of a Gaussian wave-packet centered initially at the stochastic 
region near the origin 
using the split-operator method \cite{10}. 
For the simulations presented in this letter the numerical
values of the Planck constant and 
the coupling are $\kbar= 0.29$ and $\Omega=0.65$, respectively.
To avoid numerical aliasing errors due to the periodicity of the
finite-discrete Fourier transformation, the spatial size of the
quantized grid is taken sufficiently large, $-80\le x \le 80$,
with 4096 grid points.  This allows us to resolve momenta up to
$p\simeq 23$, which is sufficient for our purpose. 
To make a comparison to the classical case we calculate 
4096 trajectories starting from 
a classical Gaussian ensemble centered initally at the origin and
having the same widths in the position and momentum 
as the quantum wavepacket.

In the top of Fig.2 we show
the spreads $\Delta{x}$ and $\Delta{p}$ in position and momentum
of the classical and quantum mechanical distributions as functions of time.
In order to remove the fast oscillations 
we have averaged the spreads over
one cycle of the $rf-$field. There are two main stages in the time dependence 
of the momentum and position spreads. In the short time behaviour,
that is for $t< 50$, there is no significant difference between the
classical (upper line) and the quantum mechanical (lower line) spreads.
In the second stage, which characterizes the long time behavior, 
that is for $t>50$, there is a considerable
difference between the classical and the quantum mechanical spreads:
whereas the classical ones increase monotonically \cite{diffusion}
the corresponding 
quantum mechanical ones oscillate with a small amplitude around an
average value. This is the first indication that the
quantum mechanical distributions show dynamical localization.  

To bring this out most clearly we show in the lower part of 
Fig.2 the time averaged probability distributions of position and momentum.
The classical distributions are broad, 
while the quantum mechanical ones are dynamically 
localized. We have also verified that this phenomenon
is not sensitive to the initial position of the wave packet in the
chaotic region and hence not destroyed by small fluctuations 
in the optical phase $\phi$.

The wave packet analysis presented above
provides a direct comparison between the quantum
and the classical distributions. In order to describe and understand
dynamical localization in a purely quantum mechanical language,
we now calculate the Floquet quasienergies $\mu_k$ 
and the corresponding eigenstates $\psi_k(x,t) = e^{-i\mu_k t} u_k(x,t)$
of the trapped ion -- laser field 
system. Here the functions $u_k(x,t)$ are periodic in time with period $\pi$.
We can obtain the Floquet states as the eigenstates 
of the eigenvalue equation
\begin{equation}
U(t+\pi , t) \psi_k(x,t) = e^{-i\mu_k \pi} \psi_k(x,t)   \label{eigsys}
\end{equation}
with the time-evolution operator $U(t+\pi , t)$
propagating the state over one time period.
We construct $U(t+\pi , t)$ by integrating the Schr\"odinger
equation for the 200 lowest eigenstates of a
stationary reference oscillator \cite{Glauber92,Fourier-analysis}.

In Fig.3 we show the quasienergies as a function of the coupling.
For zero coupling, that is for $\Omega=0$ the problem reduces to
the time-dependent harmonic oscillator with
the quasienergies $\mu_k=\mu(k+\frac{1}{2})$
ordered in a natural way with increasing
$k=0,1,...$. Here $\mu$ is the characteristic exponent of the
corresponding classical Mathieu equation. 
When the field is turned on, the two
lowest quasienergies start to grow, and become almost degenerate 
for couplings $\Omega \ge 0.2$. 
This behavior results from an effective double-well
potential caused by the standing laser field and 
the trap potential \cite{Cirac94}.
The quasienergies of this ground state doublet 
approach zero for $\Omega \sim 0.48$. They reappear from the top as
a characteristic line in the upper right corner
due to the periodicity of $\mu_k \pm 2 = \mu_k$ following 
from Eq.(\ref{eigsys}). Moreover, characteristic double lines 
connecting the lower left corner 
and the upper right corner emerge from the coupling strength axis
at $\Omega \simeq 0$, $0.1$ and $0.3$. 
These lines are associated with higher energy doublets.
We conclude this discussion of the quasienergies by emphasizing that
the existence of nearly degenerate doublets 
is connected \cite{Holthaus96} to the possibility of quantum tunneling 
between the stable islands \cite{Robinson95} 
-- a topic discussed in more detail in an upcoming paper. 

The Floquet solutions $\psi_k$ form a complete orthogonal basis.
Hence we can represent any wavepacket solution 
$\psi(x,t)$ of the Schr\"odinger equation (\ref{t_ev}) as a superposition
$\psi(x,t) = \sum_k a_k \psi_k(x,t)$
with time independent coefficients $a_k$. In this respect Floquet states 
play a role similar to that of the energy eigenstates in the 
time-independent case, and the quantities 
$\mu_k$ have the meaning of time averaged energies.
In Fig.4 we show the expansion coefficients $|a_k|^2$ for
an initial Gaussian wave packet used in the discussion of Fig.2.
We find that indeed the decomposition is 
strongly localized: Only four states
are enough to cover sixty percent of the initial
wave packet. This guarantees that the classical diffusion will stop in the 
quantum case, as shown in Fig.2.

In the insets of Fig.4 we depict the position distributions $|\psi_2|^2$ and
$|\psi_6|^2$ corresponding to a dominant and a small contribution to
the wave packet, respectively. We note that the state $\psi_2$ has major
contributions at the center of the trap.  In contrast 
$\psi_6$ has maxima at $x\simeq \pm \pi$, that is in the
neighbourhood of the stable islands. 
The mixed phase space with stochastic domains and stable islands
results from the combination of the trap potential and the standing wave, and
reflects itself in the three peak structure of
the position distribution in Fig.2. 
Note that classically these peaks are missing.
Indeed classically, a particle starting 
from the stochastic region cannot reach the 
islands. Quantum mechanically, however, this is possible through
tunneling. 

The observation of dynamical localization is possible with the present
ion trap systems. Indeed taking the experimental trap parameters from
\cite{Wineland} and considering a dipole transition from the ground
state of $^9$Be$^+$ we obtain for a driving frequency $\omega/2\pi\sim
200$MHz the values $q\sim0.2$ and $\kbar\sim0.015$.  In order to
achieve the value $\kbar\sim0.3$ used in our simulations we need a
smaller frequency such as $\omega/2\pi\sim 10$MHz; 
to keep $q$ in the stable region
the applied voltage has to be \cite{Paul} smaller, or
the trap size larger, than in \cite{Wineland}.  To be consistent with
the assumption of far detuning, the term $\Omega_0/\Delta\equiv
\epsilon$ in the dimensionless coupling
$\Omega=(\kbar\Delta)/(4\omega) \epsilon^2$ has to be small. For
$\epsilon=0.1$ and the detuning $\Delta/2\pi \sim 10$GHz we obtain
$\Omega=0.65$, as used in our simulations. 

We conclude by summarizing our main results.  The motion of a trapped
ion interacting with a laser field shows the phenomenon of dynamical
localization in position and momentum.  This phenomenon can be
observed by measuring the fluorescence light from the ion -- a
technique \cite{Moore,Wineland,Walther} which has already proven to be
an invaluable tool in studies of the quantum mechanical center-of-mass
motion of particles. Moreover we emphasize that the three-peak stucture
of the quantum mechanical position distribution, reflecting the mixed
phase space, opens up new possibilities to study the phenomenon of
dynamical quantum tunneling.

We thank M.\ G.\ Raizen and M.\ Holthaus for 
discussion and useful comments. P.\ T.\ and V.\ S.\ acknowledge 
the support of the Deutsche Forschungsgemeinschaft and 
M.\ E.\ the support of the Egyptian Government.

\begin{figure}

\caption{Poincar\'e surface of section for an ion moving in a strongly detuned 
standing wave laser field and a time dependent harmonic potential. 
The dynamics is chaotic all over the phase space except for two 
small stable islands around the 
phase space points $(x=\pi,p=0)$ and $(x=-\pi,p=0)$. The circle with its 
center at the origin
of phase space denotes the contour line of exponential decay of the Gaussian
phase space distribution of the initial wave packet. The box in the 
upper left corner shows the area of $2\pi \protect\kbar $. 
Here we have used the trap parameters $a=0.0$, $q=0.4$ 
and the coupling $\Omega=0.65$.}

\caption{Classical and quantum dynamics of a single ion moving
under the influence of a strongly detuned standing laser field and a time dependent
harmonic potential. On the top we show the time dependence of the widths of the 
classical (upper curve) and quantum mechanical (lower curve) position (left) and
momentum (right) distributions. We find that classically both widths increase with time
whereas in the quantum case they oscillate around an average value.
On the bottom we show in a semilogarithmic plot the corresponding 
position (left) and momentum (right)
distributions averaged over time in an interval of $\Delta t= 50\pi$ around $t=475\pi$.
Indeed the classical distributions are broad Gaussians 
giving rise to a quadratic curve  
whereas the quantum ones consist of narrow 
distributions which rest on a broad pedestal. 
Here we have used a wave packet of width $\sigma_{0x}^2=\protect\kbar=0.29$ and the 
trap parameters $a=0.0$, $q=0.4$ with the coupling $\Omega=0.65$.}

\caption{Quasienergies $\mu_k$ as a function of the coupling 
$\Omega$. The black diamonds
at $\Omega=0$ correspond to the seven lowest quasienergies \protect \cite{Glauber92} 
$\mu_k = \mu (k+\frac{1}{2})$ of the time-dependent
harmonic oscillator. For the trap parameters $a=0.0$, $q=0.4$ and the scaled
time $t=\omega \tilde{t}/2$ we find 
\protect \cite{Glauber92} $\mu = 0.29$. The two lowest
eigenstates become degenerate for $\Omega>0.2$ reflecting the appearance of
an effective double-well potential with growing $\Omega$. Avoided crossings
between the states of the same parity appear.
Here we have used a basis of 40 lowest eigenstates 
of the reference oscillator.}

\caption{Expansion coefficients $|a_k|^2$ of the 
initial Gaussian wavepacket of Figs.\ 1 and 2 into Floquet states $\psi_k$.
The states $\psi_1$ and $\psi_2$ cover almost half of the wavepacket.
The insets show the position distributions of the strongly contributing state 
$\psi_2$ with weight $|a_2|^2=0.21$ and the less important state $\psi_6$
with $|a_6|^2=0.03$. For comparison we represent
the initial Gaussian by a dotted line.
Whereas the state $\psi_2$ is strongly localized 
at the origin, the state $\psi_6$ has dominant
peaks in the neighbourhood of the stable islands of Fig.1.}

\end{figure}

\begin{references}
\bibitem{Anderson} P.\ W.\ Anderson, Phys.\ Rev.\ {\bf 109}, 1492 (1958).
\bibitem{Haake92}
F.\ Haake, {\it Quantum Signatures of Chaos} (Springer-Verlag, Berlin 1992).
\bibitem{review} 
For a review see for example P.\ M.\ Koch and K.\ A.\ H.\ van Leeuwen,
Phys.\ Rep.\ {\bf 255}, 289 (1995); G.\ Casati, Phys.\ Rev.\ A {\bf 45},
7670 (1992).
\bibitem{Moore} F.\ L.\ Moore et al., 
Phys.\ Rev.\ Lett.\ {\bf 73}, 2974 (1994).
\bibitem{Bialynicki-Birula94}
We note that this phenomenon is different from the formation
of Trojan wave packets in circularly polarized fields as
discussed by
I.\ Bialynicki-Birula et al.,
Phys.\ Rev.\ Lett.\  {\bf 73}, 1777 (1994); 
M.\ Kali\'nski and J.\ H.\ Eberly, Phys.\ Rev.\ Lett.\  {\bf 77}, 2420 (1996);
A.\ Buchleitner and D.\ Delande, Phys.\ Rev.\ Lett.\ {\bf 75}, 1487 (1995).
In the present problem we cannot transform the time dependence.
\bibitem{Paul}
W.\ Paul, Rev. Mod. Phys. {\bf 62}, 531 (1990).
\bibitem{Wineland} 
It is now possible to create experimentally quantum
states of the motion of an ion in the trap, see for example
D.\ M.\ Meekhof et al.,
Phys.\ Rev.\ Lett.\ {\bf 76},  1796 (1996),
and in particular Schr\"odinger cats
C.\ Monroe et al., Science 
{\bf 272}, 1131 (1996).
Also the measurement of the full density matrix is now 
possible as verified by 
D.\ Leibfried et al., submitted to Phys.\ Rev.\ Lett.\ 
\bibitem{Blumel}
In the absence of the standing wave chaos can arize from either the 
Coulomb interaction of at least two ions 
[R.\ Bl\"umel et al., Nature {\bf 334}, 309 (1988); M.\ Moore and R.\ Bl\"umel,
Phys.\ Rev. A {\bf 48}, 3082 (1993)] or a nonlinear trapping force
such as in sextapole trap.
\bibitem{Chacon94} 
For a classical discussion and treatment of this problem, see
R.\ Chac\'{o}n and J.\ I.\ Cirac, 
Phys.\ Rev.\ {\bf A 51,} 6, 4900, (1994).
\bibitem{Walther}
G.\ Birkl et al., Europhys.\ Lett.\ {\bf 27}, 197 (1994).
\bibitem{Toschek}
B.\ Appasamy et al., IQEC'96 Technical Digest (Optical Society of
America, Washington DC 1996).
\bibitem{11} M.\ Arndt et al.,
Phys.\ Rev.\ Lett.\  {\bf 67}, 2435 (1991); 
F.\ L.\ Moore et al., Phys.\ Rev.\ Lett.\  {\bf 75}, 
4598 (1995); R.\ Graham and S.\ Miyazaki, Phys.\ Rev.\ A {\bf 53},
2683 (1996).
\bibitem{Kazantsev90}
A.\ P.\ Kazantsev, G.\ I.\ Surdutovich and V.\ P.\ Yakovlev, {\it
Mechanical Action of Light on Atoms} (World Scientific, Singapore 1990).
\bibitem{phi}
Note that the case of $\phi=\pi/2$ and $\Delta> 0$ or $\phi=0$ with $\Delta< 0$
will lead to nearly integrable dynamics at the origin of the phase space.
\bibitem{10} M.\ D.\ Feit et al.,
J.\ of Comput.\ Phys.\ {\bf 47,} 412 (1982).
\bibitem{diffusion}
It is interesting to note that the classical motion shows anomalous diffusion
that is $\Delta x \sim t^{1/4}$ and $\Delta p \sim t^{1/4}$, as can be
verified from a log-log plot of the top figures of Fig.2.
\bibitem{Glauber92}
R.\ J.\ Glauber, {\it Laser manipulation of Atoms and Ions}, Proc. Int. School 
of Physics 'Enrico Fermi' Course 118, ed. E.\ Arimondo et al.\ (North Holland,
Amsterdam 1992); see also
G.\ Schrade et al., Appl.\ Phys.\ B (to be published).
\bibitem{Fourier-analysis}
Another method to find the quasienergies $\mu_k$ is to
make a Fourier analysis leading to three-term recurrence relations
between the Fourier coefficients. We have used this as an independent
check of $\mu_k$ obtained from (\ref{eigsys}).
\bibitem{Cirac94}
J.\ I.\ Cirac et al., Europhys.\ Lett.\ {\bf 27},
123 (1994).
\bibitem{Holthaus96}
H.\ P.\ Breuer and M.\ Holthaus, Ann.\ Phys.\ (N.Y.) {\bf 211}, 249 (1991).
\bibitem{Robinson95}
See also the last paragraph in J.\ C.\ Robinson et al., Phys.\ Rev.\ Lett.\ 
{\bf 74}, 3963 (1995).For a recent atom optics experiment on ordinary
tunneling in a tilted washboard potential see C.\ F.\ Brarucha et al.\, 
``Observation of Atomic Tunneling from an Accelerating Optical Potential'',
submitted to Phys.\ Rev. A. 
\end{references}
\end{document}